# Can Hubble tension be eased by invoking a finite range for gravity?

**Louise Rebecca [a], C Sivaram [b], Dominic Sebastian [a], Kenath Arun *[a]**

[a] Department of Physics and Electronics, CHRIST (Deemed to be University), Bangalore, India

[b] Indian Institute of Astrophysics, Bangalore, India

**Abstract:** The estimation of the Hubble constant in the past few decades has increasingly become more accurate with the advance of new techniques. But its value seems to depend on the epoch at which the measurements are made. The Planck estimate of the Hubble constant from the observations of the cosmic microwave background radiation in the early universe is about $67 \, km\backslash s\backslash Mpc$, whereas that obtained from the distance indicators at the current epoch is $\sim 73 - 74 \, km\backslash s\backslash Mpc$. This discrepancy between the two groups of measurement is termed as the Hubble tension which has gained much attention in the past few decades with growing significance as measurements from both, the early and the late universe, studies continue to produce results with increasing precision. In this work, we propose a modification to gravity by considering a finite range gravitational field as an alternate explanation for this discrepancy in the value of the Hubble constant.

*Corresponding author:

Louise Rebecca: e-mail: louiserheanna.rebecca@res.christuniversity.in

C Sivaram: e-mail: sivaram@iiap.res.in

Dominic Sebastian: e-mail: dominic.sebastian@phy.christuniversity.in

Kenath Arun: e-mail: kenath.arun@christuniversity.in; https://orcid.org/0000-0002-2183-9425

# 1. Introduction

One of the most significant discoveries of the twentieth century is the expansion of the Universe by Edwin Hubble in 1929 (Hubble 1929). This discovery established a linear relationship between the apparent distances to galaxies and their recessional velocities. Friedman in 1922 derived a set of equations for an expanding homogeneous and isotropic universe from Einstein's field equations (Friedman 1922). The equations for negative curvature were later derived in 1924 (Friedman 1924). Lemaitre (Lemaitre 1927) provided a mathematical solution for an expanding Universe with general relativity which explained the observed receding velocities of galaxies. For any observer, galaxies at greater and greater distances recede with larger and larger velocities following the relation,

$$v = H_0 d \qquad (1)$$

Equation (1) is known as the Hubble-Lemaitre law, where $v$ is the recessional velocity, $H_0$ is the Hubble constant and $d$ is proper distance to the galaxy from the observer. The recession velocities can be derived from the galaxy's redshift. But, to measure the distances one needs objects of known intrinsic luminosity (standard candles).

One method of estimation of the Hubble constant is by observation of anisotropies and other features of the cosmic microwave background radiation, which provides a snapshot of the early universe. Another process involves estimation of the Hubble constant 'locally' by determining the distances to nearby galaxies through their distance-redshift relation. These methods involve calibration of luminosities of standard candles such as Cepheid variables, type Ia supernovae, tip of red giant branch (TRGB), etc. The late universe measurements give a value of Hubble constant that is significantly different from that measured from the CMBR (Baxter and Sherwin 2021; Mukherjee et al. 2020; Soltis, Casertano and Riess 2020; Riess et al. 2020; Dutta et al. 2019; Freedman et al. 2019; Alberto et al. 2019; Yuan et al. 2019; Freedman et al. 2020). The discrepancy in the Hubble parameter is present irrespective of the quantity of bins used to split the Pantheon sample, the amount of parameters that may be changed, or, more significantly, the existence of an additional cosmological probe (BAO) even binned (Di Valentino et al. 2021; Dainotti et al. 2022). Therefore, it is evident from these observational results that this discrepancy in the measured value of Hubble constant may not be due to observational errors. Hence, one may have to look for a theoretical framework to explain this apparent inconsistency.

Theories involving modifications to the standard ΛCDM model including early dark energy, late dark energy, dark energy models with six degrees of freedom, models with extra interactions, unified cosmologies, etc. have been proposed to solve the problem of Hubble tension. (di Valentino et al. 2021; Knox and Millea 2020; Beenakker and Venhoek 2021). Another widely argued resolution involves applying changes to the late time expansion theory. (Lemos et al. 2019; Benevento et al. 2020; Camarena and Marra 2021). But it was recently shown that with a 'dynamics free' inverse distance ladder that changes to late time physics cannot solve the Hubble tension problem (Efstathiou, 2021). Pantheon sample shows decreasing trend of $H_0$ with redshift even after dividing the Pantheon sample into number of bins and allowing the variation of parameters (Dainotti et al., 2022). One possible solution could be varying Einstein constant governed by a slow evolution of a scalar field which mediates the gravity-matter interaction. But the theory did not produce the independence of $H_0$ on redshift at any redshift bin. (Dainotti et al., 2023).

**2. Hubble Constant Measurements**

The very first measurement of the Hubble constant ($H_0$) – by Hubble himself – used Cepheid variables as standard candles yielding a value of about $500 \ km\backslash s\backslash Mpc$ (Hubble 1929). This value is found to be more than five times the value obtained by subsequent measurements over the past several decades. The Hubble Space Telescope's key project established the most precise determination of $H_0$ with secondary distance measurements over a range of $60 - 400 \ Mpc$ with Cephids as standard candles. The project yielded a $H_0$ value of $72 \pm 8 \ km\backslash s\backslash Mpc$.

There has been further progress in reducing known systematic errors in such measurements with the increase in the samples of Type Ia supernovae (Riess et al. 2019; Pietrzynski et al. 2019; Riess et al. 2018a; 2018b). The Wide Field Camera 3 on the Hubble Space Telescope was used to reduce the uncertainty from 3.3 – 2.4 per cent. Most of this improvement comes from recent near-infrared observations of Cepheid variables in eleven galaxies hosting recent type Ia supernovae. This observation more than doubled the sample of reliable type Ia to a total of 19. From this, the best estimate of Hubble constant is found to be $73.24 \pm 1.74 \ km\backslash s\backslash Mpc$ (Riess et al. 2019). New measurements (Chen et al. 2019) involving the late universe reinforced the Hubble value $H_0$ obtained from the results of Supernova for the Equation of State (SH0ES) project (Freedman et al. 2019; Birrer et al. 2018). The late universe measurements have also improved distance estimates with the usage of detached eclipsing binaries, water masers, etc., all of which produce values of $H_0$ in the range of $72 -$ 73 $km\backslash s\backslash Mpc$ (Reid et al. 2019). Results from the Megamaser Cosmology Project produced a $H_0$ of 73.9 $km\backslash s\backslash Mpc$.

The standard-candle method is not the only method that can be used to obtain $H_0$. A series of high-resolution images of the Cosmic Microwave Background Radiation was captured by the Wilkinson Microwave Anisotropy Probe (WMAP) (Jarosik et al. 2011) as an independent measure of the rate of expansion of the 'early universe'. While initial results seemed to be almost in agreement with that obtained from the 'late universe', this agreement is seen to disappear with increased refinements in the method. The nine-year data release from WMAP produced a $H_0$ of $69.32 \pm 0.80$ $km\backslash s\backslash Mpc$ (Bennett et al. 2013). This was followed by the Planck Mission (Ade et al. 2016; Aghanim et al. 2020) that acquired the highest resolution maps to-date in microwaves. Analysis of fluctuations in temperature and polarization maps lead to an excellent agreement with the standard model of cosmology. Fitting the angular fluctuations in the Planck data to the ΛCDM model results in a $H_0$ value of $67.8 \pm 0.9$ $km\backslash s\backslash Mpc$. The Baryonic Accosting Oscillations method (BAO) in collaboration with Baryon Oscillations Spectroscopy Survey (BOSS) produced a value of a $H_0$ of $67.3 \pm 0.9$ $km\backslash s\backslash Mpc$ independent of CMB anisotropies (Ryan, Chen and Ratra 2019; Aubourg et al. 2015; Greib et al. 2017). Recent measurements independent of both, the distance ladder and CMB, such as the $\gamma$-ray attenuation measurements from Fermi-LAT (Domínguez et al. 2019) produced a value of $66.6 \pm 1.6$ $km\backslash s\backslash Mpc$ (Brout et al. 2022).

Another independent measurement used alternate distance ladders such the Tip of the Red Giant Branch (TRGB) rather than Type Ia supernova producing a $H_0$ value of $69.8 \pm 0.8$ $km\backslash s\backslash Mpc$ (Freedman et al. 2019), midway between the values from Planck and SH0ES. The most recent calibration of Type Ia supernova based on the surface brightness fluctuations (SBF) method consisting of 24 supernovae hosted in galaxies that have SBF distance measurements estimated a value of $70.50 \pm 3.38$ $km\backslash s\backslash Mpc$ (Khetan et al. 2021). This is in good agreement with that obtained from the TRGB calibration.

Recent work has explored the possibility of utilizing Gama Ray Burts (GRB) and quasi-stellar objects (QSO) as standard candles. GRBs and QSOs extend the Hubble diagram ($z = 9.4$ and $z = 7.642$ respectively) far beyond the limit of SNe Ia ($z = 2.2$), thus providing distances to events that occurred a few hundred million years after the Big Bang (Dainotti et al., 2022; Dainotti et al., 2023).

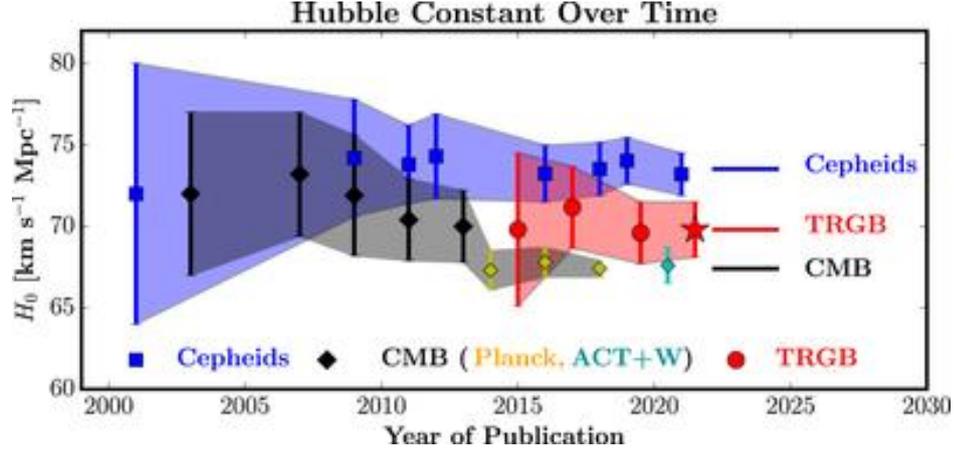

**Figure 1**: Evolution of the Hubble tension in the last 20 years using direct measurements from the distance ladder and CMB-based measurements (Freedman 2022)

As Hubble constant measures the rate of expansion of the universe, the above values suggest that the universe is expanding faster at present than at the time of the CMBR epoch. Sources of systematic error could only increase the value of $H_0$ from Planck value by not more than $1 \; km\backslash s\backslash Mpc$ under the ΛCDM model. Thus, Hubble tension, even if reduced would probably continue to exist.

## 3. Finite Range Gravity (FRG)

Although general relativity is well established through several observations, certain cosmological phenomena such as the accelerated expansion, missing matter (dark matter) and the recent tension in the Hubble constant measurements probably necessitates a need for some modifications to the theory. One such modification can be done by considering gravity to be of a finite range. There have been suggestions of a finite range gravity earlier. One such suggestion involved specific mass terms (masses of spin-2 and spin-0 fields) in addition to the field-theoretical analogue of the usual Hilbert-Einstein Lagrangian that allows the cosmological scale factor to exhibit an accelerated expansion (Babak and Grishchuk 2003). There have also been suggestions of multidimensional higher derivative theories of gravity that is marked by a non-polynomial form factor which averts extra degrees of freedom. In odd dimensions it is seen that gravity turns out to be 'finite' (Modesto 2013). Finite range gravity (short range) was also suggested in another context (Sivaram and Sinha 1976; Lord, Sinha and Sivaram 1974). For a review see for e.g. (Sivaram and Sinha 1979).

The tension in the Hubble constant between different cosmological epochs – if not due to errors in measurement – may be an actual indication of the present expansion rate of the Universe. Here we consider gravity to be of a finite range that could possibly account for an increased expansion rate in the present epoch.

Such a finite range for the gravitational field can be understood from scale invariance being broken. Scale invariance breaking (as determined by a scale like the cosmological constant, i.e. curvature) can lead to a finite range gravity, or interaction

$$S = \frac{1}{2}\int d^4x \sqrt{-g}\left[g^{\mu\nu}\phi_{,\mu}\phi_{,\nu} + \frac{1}{6}\mathcal{R}\phi^2\right] \quad (2)$$

here $\mathcal{R}$ is the curvature scalar. As scale invariance is preserved, $\phi$ is a massless scalar. Scale invariance of the above can be easily seen under scale transformations,

$$g_{\mu\nu} = \bar{g}_{\mu\nu}\phi^{-2} \quad (3)$$

$$\sqrt{-g} = \sqrt{-\bar{g}}\phi^{-4} \quad (4)$$

Hence we have,

$$\sqrt{-g}\,g^{\mu\nu}\phi_{,\mu}\phi_{,\nu} = \phi^{-2}\phi_{,\mu}\phi_{,\nu}\bar{g}_{\mu\nu}\sqrt{-\bar{g}} \quad (5)$$

$$\frac{1}{6}\mathcal{R}(g)\phi^2\sqrt{-g} = \frac{1}{6}\sqrt{-\bar{g}}\mathcal{R}(\bar{g}) - \sqrt{-\bar{g}}\,g^{\mu\nu}\phi_{,\mu}\phi_{,\nu}\phi^{-2} \quad (6)$$

So that,

$$\frac{1}{2}\int d^4x \sqrt{-g}\left[g^{\mu\nu}\phi_{,\mu}\phi_{,\nu} + \frac{1}{6}\mathcal{R}\phi^2\right] \quad (7)$$

is just Einstein action, with an effective gravitational constant,

$$G_{\mu\nu} = 6\phi^2\left[T_{\mu\nu}(\phi) + T_{\mu\nu}(\theta)\right] \quad (8)$$

This gives,

$$\Phi\left[\Box - \frac{1}{6}\mathcal{R}\right] = T_\alpha^\alpha(\theta) \quad (9)$$

So, $T_\alpha^\alpha(\theta) = 0$ is traceless and arising from scale invariance.

The scale invariance must be broken to allow for massive system, i.e.

$$T_\alpha^\alpha(\theta) \neq 0 \quad (10)$$

This can be done by adding a mass term,

$$I_{mass} = \frac{1}{2}m^2 \int d^4x \sqrt{-g}\,\phi^2 \quad (11)$$

This implies, $m^2\phi^2 = T_\alpha^\alpha(\theta)$ \quad (12)

Where $\phi$ would have a range of $\frac{\hbar}{mc}$

And, $T_\alpha^\alpha \sim \frac{m^4 c^4}{4\pi\hbar^3}$ \quad (13)

So,

$$\phi^{-2} \cong \left(\frac{\hbar}{mc}\right)^{-2}(T_\alpha^\alpha)^{-1} \quad (14)$$

$$\phi^{-2} \cong \left(\frac{\hbar}{mc}\right)^{-2} \frac{m^4 c^4}{4\pi\hbar^3} = \frac{4\pi\hbar}{m^2 c^2} \sim \frac{8\pi G}{c^4} \tag{15}$$

Here,

$$\left(\frac{mc}{\hbar}\right)^2 \approx 10^{-56} cm^{-2} \; (\sim \Lambda) \tag{16}$$

I.e., an effective cosmological constant term arising from the mass term.

So, $\frac{mc}{\hbar} \approx \frac{1}{\sqrt{\Lambda}} (\approx 10^{28} cm)$ fixes the range of the field.

The mass term becomes important when

$$r \sim (\Lambda)^{-1/2} \approx 10^{28} cm \sim \left(\frac{mc}{\hbar}\right) \tag{17}$$

This breaking of scale invariance is analogous to a phase transition, mass $m$ having a finite value, in the case where the scale invariance is broken

$$T_0^0 = \frac{1}{8\pi}\left[\left(\frac{\partial\phi}{\partial r}\right)^2 + \mathcal{R}^2 \phi^2\right] g_{00} \tag{18}$$

For massless field,

$$T_{00} = \frac{1}{2}\partial_\mu \phi \partial_\nu \phi = \frac{1}{2}\left(\frac{\partial\phi}{\partial r}\right)^2 \tag{19}$$

$$T_{11} = \frac{1}{8\pi}\left[\left(\frac{\partial\phi}{\partial r}\right)^2 + \mathcal{R}^2 \phi^2\right] g_{11} \tag{20}$$

(Following the usual prescription)

$$T_{\mu\nu} = \overline{\partial_{\mu\nu}} L - \frac{\partial L}{\partial g_{\mu\nu}} \tag{21}$$

where $L$ is the Lagrangian.

With this finite-range model, the gravitational gets potential is modified (solving equation (18) for a massive field) as

$$V = -\frac{GM}{r} e^{-\frac{r}{r_0}} \tag{22}$$

where, $r_0$ is a constant indicating the range of the force, $r$ is a scale factor.

For $r \ll r_0$, the potential reduces to the usual Newtonian potential, i.e., $V = -\frac{GM}{r}$. The modification becomes relevant when $r$ is comparable to $r_0$. Gravity will weaken following equation (22) as $r$ becomes comparably larger than $r_0$. Where, $r_0$ is related to the scale fixed in the early universe (broken scale invariance).

The massive field thus modifies the energy density as $\frac{8\pi G\rho}{3} e^{-\frac{r}{r_0}}$ in the usual Friedmann field equations.

The idea is that the scale invariance is broken in the early universe through a scale fixed by cosmological constant, and this gives rise to a minimum background curvature. This term however is not important in the early universe, but at the present epoch when the Hubble radius becomes comparable to the scale (fixed by the cosmological constant) now becomes significant. This would modify the Friedman equation as shown in next section.

**4. Deduction of Friedmann equation**

The time evolution of a homogenous, isotropic Universe can be deduced from Newtonian dynamics. The total energy of the Universe can be written as the sum of the kinetic energy and the potential energy, i.e.

$$U = T + V \qquad (23)$$

where, $T \left(= \frac{1}{2} H^2 r^2\right)$ is the kinetic energy per unit mass and $V \left(= -\frac{4}{3} \pi G \rho r^2\right)$ is the potential per unit mass. For a flat Universe (zero curvature) the total energy is zero. Therefore, we have,

$$\frac{1}{2} H^2 r^2 = \frac{4}{3} \pi G \rho r^2 \qquad (24)$$

By modifying the potential term (for the finite range) as mentioned in section 3 to equation (24) we get,

$$\frac{1}{2} H^2 r^2 = \frac{4}{3} \pi G \rho r^2 \left(1 + e^{-\frac{r}{r_0}}\right) \qquad (25)$$

It is well known that, the Newtonian result given by equation (24) is just the same as what we get in GR, by solving the Friedmann metric with an energy momentum tensor for pressureless matter $(P = 0)$, i.e.,

$$H^2 = \frac{\dot{r}^2}{r^2} = \frac{8 \pi G \rho}{3} + \frac{\Lambda c^2}{3} \qquad (26)$$

where, $r$ is the scale factor, $\dot{r}^2$ is the kinetic energy per unit mass, $\frac{\dot{r}^2}{r^2}$ is the Hubble expansion parameter, and $\rho$ is the matter density. As usual, the curvature $k$ is taken to be zero. We retain the cosmological constant term (Arun, Gudennavar and Sivaram 2017).

With the modification as detailed in section 3 applied to the potential energy term, equation (26) now takes the form,

$$H'^2 = \frac{\dot{r}^2}{r^2} = \frac{8 \pi G \rho}{3} \left(1 + e^{-\frac{r}{r_0}}\right) + \frac{\Lambda c^2}{3} \qquad (27)$$

Considering the modified potential energy term, equation (27) follows from the same Newtonian consideration. Curvature is neglected $(k = 0)$ and spherical symmetry of the potential is retained, so the isotropy and homogeneity are retained. There is no anisotropic expansion as the potential is spherically symmetric.

The new potential energy per unit mass term in equation (25) chiefly changes the expansion rate by $e^{-\frac{r}{r_0}}$ as the Universe undergoes expansion from $r_0$ to $r$. When $r \ll r_0$, i.e., in the early Universe when the modified term is not relevant, equation (27) takes the usual form as in equation (26). The basic argument is that $H^2 - \frac{8\pi G\rho}{3}$ is modified to $H'^2 - \frac{8\pi G\rho}{3}\left(1 + e^{-\frac{r}{r_0}}\right)$. As gravity weakens due to its finite range, to counteract the reduced potential energy term, the kinetic energy term now increasing, hence resulting in $H'^2 > H^2$.

The Hubble parameter value from equation (27) as the Universe expanded from $r_0$ (at the corresponding redshift, $z$ to the present size $r$ is plotted in figure 2 and it is compared with the existing data obtained from literature (Burns et al. 2018, Bailes et al. 2021, Sedgwick et al. 2021, Wang et al. 2023, Chen et al. 2019).

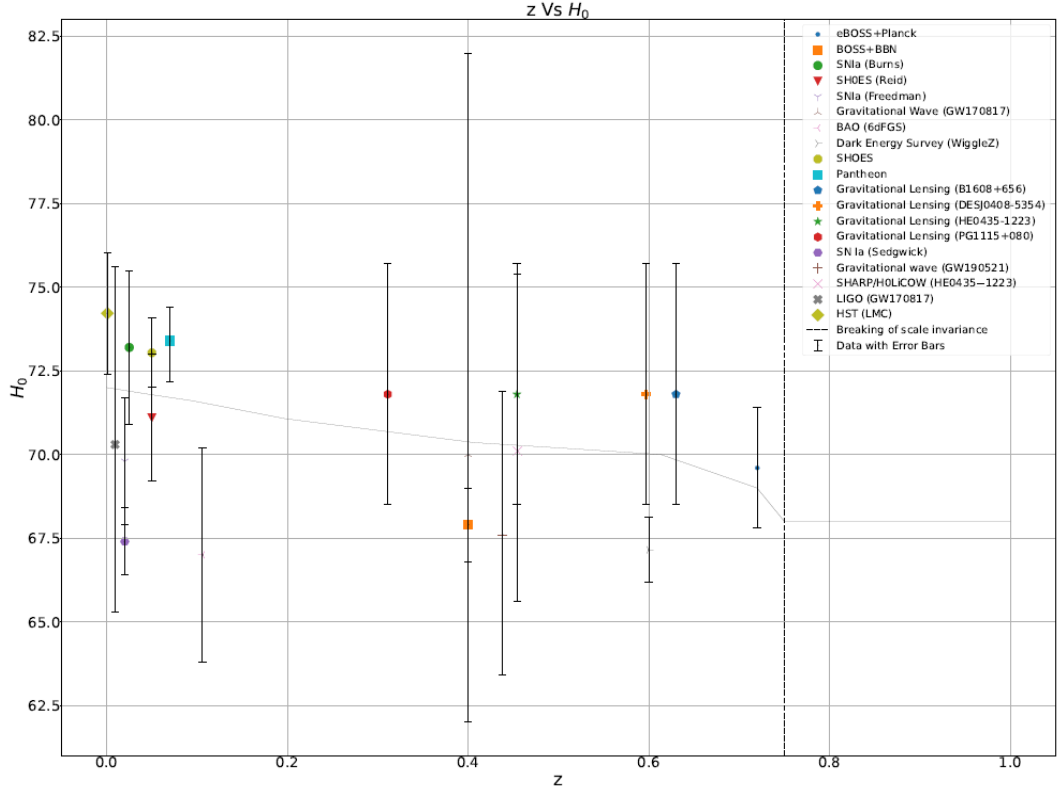

**Figure 2**: Hubble parameter versus *z*. Our model (solid line) compared to observations.

For the present epoch, with $r = 10^{28} cm$, for a decrease in the value of the second term by 10 per cent (i.e., $e^{-\frac{r}{r_0}} \approx 0.1$), the value of $r_0$ has to be $\approx 5 \times 10^{27} cm$. This redshift that corresponds to the breaking of scale invariance as indicated by the vertical dotted line in figure

2. This range for the gravitational field corresponds to a graviton mass which may be related to the cosmological constant through

$$m_{grav} = \frac{h}{c}\sqrt{\Lambda} \approx 2 \times 10^{-65} g \qquad (28)$$

This thereby implies that for $r > r_0$ gravity weakens and $H^2$ can become 10 per cent larger indicating an increase in the rate of expansion. This could probably account for a higher value of Hubble constant as seen at the present epoch.

## 5. Conclusion

Considering the fact that the observed inconsistency in Hubble constant measurements may not be due to observational errors, we need to look for alternate models to account for the increased value of the Hubble parameter, i.e., to reduce this tension. In an earlier work we invoked a modification of the gravitational field on large scales (low accelerations) that could provide a possible solution for this discrepancy and perhaps help in relaxing the Hubble tension (Sivaram, Arun and Rebecca 2021).

In this work we extend the idea through a modification to gravity by invoking a finite range gravitational field to explain the discrepancy. Finite gravity has been introduced in other contexts earlier. Here the introduction of an exponential term to the gravitational potential modifies the Friedmann equation and this new term changes the expansion rate as the Universe expands beyond a critical size. The weakening of the potential energy term due to finite range correspondingly increases the kinetic term i.e., the expansion rate. We also suggest that the gravitational field's range corresponds to a graviton mass and this may be related to the cosmological constant. With this modified gravity we can account for the ~10% change in the Hubble constant measured 'locally' versus that measured at early epochs.

Data Availability Statement: No Data associated in the manuscript